\documentclass[a4paper,noindent]{JHEP3}

\usepackage{epsfig, subfigure}
\usepackage{graphicx}
\usepackage{amsmath}
\usepackage{amssymb}
\usepackage{bbm}
\usepackage{epsfig}
\usepackage{slashed}

\def\ga{\mathrel{\raise.3ex\hbox{$>$\kern-.75em\lower1ex\hbox{$\sim$}}}}
\def\la{\mathrel{\raise.3ex\hbox{$<$\kern-.75em\lower1ex\hbox{$\sim$}}}}

\newcommand{\lam}{\lambda}
\def\lsim{\mathrel{\rlap{\lower4pt\hbox{\hskip1pt$\sim$}}
    \raise1pt\hbox{$<$}}}                % less than or approx. symbol
\def\gsim{\mathrel{\rlap{\lower4pt\hbox{\hskip1pt$\sim$}}
    \raise1pt\hbox{$>$}}}                % greater than or approx. symbol

\usepackage{cite}

\makeatletter
\renewcommand\appendix{\par
  \setcounter{section}{0}%
  \setcounter{subsection}{0}%
  \renewcommand\thesection{\@Alph\c@section}
  \setcounter{figure}{0}%
  \setcounter{table}{0}%
  \renewcommand\thefigure{\thesection.\@arabic\c@figure}
  \renewcommand\thetable{\thesection.\@arabic\c@table}
  }
\makeatother

\title{Higgs boson phenomenology in $\tau^+ \tau^-$ final states at the LHC}

\author{Alexander Belyaev$^{1,2}$, Renato Guedes$^{1}$,
Stefano Moretti$^{1,2,3}$,
and Rui Santos$^{1,2}$\\
{$^1$ NExT Institute and School of Physics and
Astronomy, University of Southampton Highfield, Southampton SO17
1BJ, UK.}\\
{$^2$ Particle Physics Department, Rutherford Appleton Laboratory, Chilton, Didcot, Oxon OX11
0QX, UK.}\\
{$^3$Dipartimento di Fisica Teorica, Universit\`a di Torino,
Via Pietro Giuria 1, 10125 Torino, Italy.}

\\ E-mails: \email{a.belyaev@soton.ac.uk}, \email{r.b.guedes@soton.ac.uk}
\email{stefano@phys.soton.ac.uk},
\email{r.a.santos@soton.ac.uk}
}

\abstract{We perform a detailed parton level study on the feasibility of the
detection of a Higgs boson
in the gluon fusion process $pp (gg+gq) \to h + jet \to
\tau^+ \tau^- + jet$  at the Large Hadron Collider (LHC) for $\sqrt{s}=14$ $TeV$.  The obtained
results are applied to a few chosen Beyond the Standard Model (BSM)
scenarios where the branching ratio of a Higgs boson decaying into
a $\tau^+ \tau^-$ pair is enhanced
as compared to the Standard Model (SM) case. We present
the parameter space of the BSM
scenarios that can be observed at the LHC
and conclude that some regions of the parameter space can be  probed with just a few $fb^{-1}$
of  integrated luminosity. Noticeably, our results are
presented in a form which potentially allows their application to any generic model
giving rise to a $pp (gg+gq) \to h + jet \to \tau^+ \tau^- + jet$
signature.
}

\preprint{SHEP-09-20 \\ DFTT-59-2009}

\begin{document}
%------------------------------------------------
\section{Introduction}
\label{sec:intr}

One of the main aims of the LHC is the unraveling of the mechanism of Electro-Weak Symmetry Breaking (EWSB). In particular, the discovery of Higgs boson(s) would confirm one of the several possible scenarios of EWSB. Presently there are  no hints about the underlying scalar sector.  Searches~\cite{CDFHiggs, D0Higgs} and detailed studies~\cite{Aad:2009wy, Abdullin:2005yn, Cavalli:2002vs} devoted to a light ($O(100)$ $GeV$)  CP-even Higgs boson were performed for the SM and for the Minimal Supersymmetric Standard Model (MSSM)~\cite{MSSMHiggs}. Among the several production mechanisms of a SM Higgs boson, $h$, at the LHC, the most important ones are gluon
fusion, $gg \to h$ (GGH)~\cite{Georgi:1977gs}, and  vector boson fusion~(VBF)~\cite{Rainwater:1998kj}, $q q' \to q q'h $. Although a light  Higgs boson decays predominantly to $b \bar{b}$, the huge QCD background makes it virtually impossible to detect it in this decay mode. Therefore, the second most important Higgs boson decay channel, $h \to \tau^+ \tau^-$, comes into play.

The ATLAS~\cite{Aad:2009wy} and CMS~\cite{Abdullin:2005yn} collaborations have both explored  the possibility of finding a Higgs boson in final states with tau leptons at the LHC in the VBF channel. Unfortunately, the radiative $gg (gq)  \to hj \to \tau ^+ \tau^- j$ process (hereafter, $j=jet$), that is, gluon fusion in presence of an additional resolved jet, proposed as a Higgs search channel more than 20 years ago~\cite{Ellis:1987xu} did not receive much attention ever since. There is only one recent detailed study for the SM $gg \to hj \to \tau^+ \tau^- j$  process~\cite{Mellado:2004tj}. However, in~\cite{Mellado:2004tj} this process is studied together with VBF and the selection cuts are chosen so that Higgs boson production in  GGH is suppressed, in particular due to a central jet veto with $|\eta| < 1$, the latter favouring VBF.

The calculation of the $gg (gq)  \to hj$ cross section, performed in~\cite{Ellis:1987xu, Baur:1989cm}, includes the LO $p_T$ distributions of the Higgs boson with the full quark mass dependence (a detailed study of Higgs production associated with a high $p_T$ jet for the MSSM can be found in~\cite{Brein}). The process is part of the real NLO corrections to the total gluon fusion cross section which are only known in the heavy quark limit (see~\cite{Nath:2010zj} for a discussion). Therefore the corrections are only valid for small and moderate Higgs masses and $p_T$~\cite{higgsptnlo}. In this limit, NLL~\cite{higgsptnll} and NNLL~\cite{higgsptnnll} soft gluon ressumation are also available, which in turn means that there is now a reliable description of the low $p_T$ region.
It was also shown in ~\cite{Ellis:1987xu, Baur:1989cm} that the effective interaction $G_{\mu \nu}^A \, G^{\mu \nu}_A \, H$, which is equivalent to taking $m_t \to \infty$ in the complete calculation, is a good approximation provided the Higgs mass is below the top-quark mass. The Higgs $p_T$ distribution was shown to be reliable up to values of the order of 300 to 400 GeV. Finally we should note that the very low $p_T$ region is not again well described by the effective vertex because in that region the process $gg \to H$ becomes important and so are the soft radiation corrections to this same process.

This paper is devoted exclusively to a detailed study of the $gg (gq)  \to hj \to \tau^+ \tau^- j$ process at the LHC with $\sqrt{s} = 14$ TeV, for which we optimise the respective kinematical
selection criteria in the same vein as in an analogous study for the Tevatron~\cite{Belyaev:2002zz}. One of the features of this channel is that it is sensitive to models where  Higgs boson couplings to fermions are enhanced as compared to the SM and does not depend on the Higgs boson couplings to vector bosons~\cite{Belyaev:2005ct}.
The GGH process thus boasts the very important feature of being complementary to the VBF process -- a clear identification of a given EWSB pattern can only be accomplished with independent measurements of the Higgs couplings to gauge bosons and fermions, respectively.

Therefore, the process under study is quite  appealing  for SM Higgs boson searches. Moreover, the $pp \to h + jet \to \tau^+ \tau^- + jet$ process  could become much more important in BSM models such as Supersymmetry, Technicolour~\cite{Belyaev:2005ct,us}  or some specific 2HDM models where the signal rate  of Higgs production or decay  (or both) is enhanced~\cite{Belyaev:2009nh}. In this particular study, we have applied the results to models where  the Higgs decay to leptons is enhanced relative to the SM. All such models have in common the fact they have a specific type of 2HDM as submodel for EWSB. Recently, a number of these scenarios have been discussed in the literature~\cite{Aoki:2008av, Aoki:2009vf, Goh:2009wg}, wherein the Higgs Branching Ratio (BR) into leptons is enhanced relative to the SM case and can therefore effectively be probed with the process under study. These models provide Dark Matter (DM) candidates and can accommodate neutrino oscillations and a baryon asymmetry while being in agreement with all  experimental data. Earlier versions of such models were discussed in~\cite{vac1,Barbieri:2006dq}.

The plan of this paper is as follows. In Sect.~\ref{sec:signal} we discuss the Higgs signal and corresponding backgrounds in the SM.  In Sect.~\ref{sec:2HDM} we introduce 2HDMs  where leptonic Higgs decays can be enhanced and discuss their phenomenology at the LHC. In Sect.~\ref{sec:bounds} we review our results in the light of all available experimental constraints. Finally, we conclude in Sect.~\ref{sec:conclusions}.

%----------------------------------------
\section{Signal and backgrounds in the Standard Model}
\label{sec:signal}

\begin{figure}[h!]
  \begin{center}
    \epsfig{file=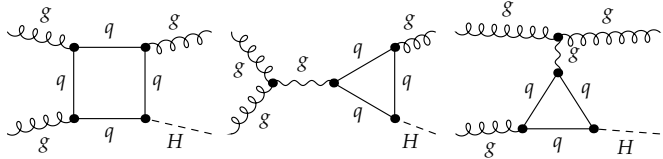,width=10cm}
%\vspace{-1.cm}
    \caption{Generic diagrams for the $gg \to gh$ process, where $q$ stands for a generic quark.}
    \label{fig:fig_diagrams}
  \end{center}
\end{figure}

The production process we are considering, $pp \to gg (q) \to hg (q) \to \tau^+ \tau^- g (q)$ proceeds, at the parton level, via the diagrams shown in Fig.~\ref{fig:fig_diagrams}. Due to the high gluon luminosity, the GGH process (of order $\alpha_S^3 \alpha_W$), although being one-loop induced, dominates the VBF one (which is of order $\alpha_W^3$). For a SM Higgs boson with mass of 120 GeV the Leading Order (LO) cross section for GGH is 15.2 pb while for VBF is only 5.2 pb. In the loop process, besides $gg \to hg $ as shown in Fig.~\ref{fig:fig_diagrams}, we have also included the $gq \to hq$ process which is approximately 20 \% of the total cross section. The remaining contribution, $q \bar{q} \to hg$ was shown to be negligible~\cite{Abdullin:1998er} and therefore was not taken into account in our study. Before starting the analysis we should point out that this is a parton level study. Effects of initial and final state radiation as well as hadronisation were not taken into account.

%%--------------------------------------------------------------
%--------------------------------------------------------------
%

The SM signal and all background processes were generated using CalcHEP~\cite{Pukhov:2004ca} where the effective vertex $ggh$ is implemented. The results were cross checked with MadGraph/MadEvent~\cite{Alwall:2007st}. Higgs BRs to $\tau^+ \tau^-$ have been evaluated with the HDECAY~\cite{Djouadi:1997yw} package. The one-loop amplitudes for the $pp \to gg (q) \to hg (q) \to \tau^+ \tau^- g (q)$ process in the 2HDM were generated and calculated with the packages FeynArts~\cite{feynarts} and FormCalc~\cite{formcalc}. The scalar integrals were evaluated with  LoopTools~\cite{looptools}. We have used the CTEQ6L parton distribution functions \cite{cteq}. The following equal  factorisation and renormalisation scales relevant for each process were chosen:  $m_h$ for the signal, $M_Z$ for $Zj$, $2 M_W$ for $WWj$, $m_{top}$ for $t\bar{t}$, $M_W$ for $Wjj$ and $\sqrt{\hat{s}}$ for $jjj$. These scales are motivated by a particular energy scale specific to each process. The jet (leptons) energies were smeared according to the following Gaussian distribution
\begin{equation}
\frac{\Delta E}{E}=\frac{0.5 (0.15)}{\sqrt{E}} \, GeV,
\end{equation}
to take into account the respective detector energy resolution effects, where $0.5$ is the factor for jets while $0.15$ is the corresponding factor for leptons.

We could a priori consider all possible $\tau$-decay modes: both taus decaying hadronically, BR$(\tau \to j)^2 = 0.65^2\simeq 0.42$; one tau decaying hadronically and the other leptonically with BR$(\tau \to j)\times $BR$(\tau \to l) = 2\times 0.65\times 0.35 \simeq 0.46$, where $l=e,\mu$; both taus decaying into leptons with BR$(\tau \to l)^2 = 0.35\times 0.35 \simeq 0.12$. With the available trigger set, the first scenario would be the less efficient although triggers for tau and missing energy were used by ATLAS in~\cite{Aad:2009wy}.

The study of purely hadronic final states is quite challenging because of its lower trigger efficiency as well as its complexity related to the identification of the three jet final state pattern. The  two remaining cases have robust trigger signatures - the events are selected by an isolated electron with $p_T^e > 22 \, GeV$  or an isolated muon with $p_T^\mu > 20 \, GeV$. Therefore, we have decided to perform the two analyses, the one where the two taus decay leptonically ($ll$) and the other one where one tau decays leptonically  and the other hadronically ($lj$). In both analyses we have taken into account the main source of irreducible background: $pp \to Z/\gamma^* j \to l l j$  for $ll$ and $pp \to Z/\gamma^* j \to l j j$, where one jet originates from a tau, for the $l j$ case. In $pp \to Z/\gamma^* j \to l l j$ we include all possible combinations of $l=e,\mu$ and in $pp \to Z/\gamma^* j \to l j j$ only the intermediate state $\tau^+ \tau^- j$ is included - the $jjj$ final state, where a jet would fake a lepton with a given probability, is taken into account in the $jjj$ background.

The main source of reducible background for the $ll$ analysis comes from $pp\to W^+ W^- j$ while for the $lj$ case it is the process $pp\to W j j$ that dominates. Process $pp\to W^+ W^- j$ is not relevant for the $lj$ analysis because while in $ll$ all possible combinations of $W$ bosons decaying to leptons contribute to the background, in the $lj$ case just one lepton can be $e, \, \mu$ or $\tau$. Moreover, if one $W$ decays into jets, the probability of a jet faking a hadronic tau makes this contribution negligible. The $pp\to W j j$ noise is larger in the $lj$ analysis because the probability of a jet faking a hadronic tau is higher than the one of a jet faking a lepton. The tau reconstruction efficiency was taken to be 0.3 and accordingly we have used a tau rejection factor against jets as a function of the jet $p_T$ using the values presented in the ATLAS study in~\cite{Aad:2009wy}. Finally, we have included the $pp \to t\bar{t}$ background taken at NLO. Two steps of the analysis allow us to discard most of the $t\bar{t}$ background: vetoing the events if the tagging jet is consistent with a $b$-jet hypothesis for $|\eta| < 2.5$ and the fact that most of this background consists of final states with two jets - if one additional jet is detected, the event is discarded. The $t\bar{t}$ background is larger for $ll$ as there are more possible combinations when the $W$ bosons decay leptonically.

%%%%%%%%%%%%%%%%%%%%%%%%%%%%%%%%%%%%%%%%%%%%%%%%%%%%%%%%%%%%%%%%%%%%%%%%%%%%%%%%%%
%\begin{widetext}
\begin{figure}[h]
\begin{center}
\includegraphics[width=7.5cm]{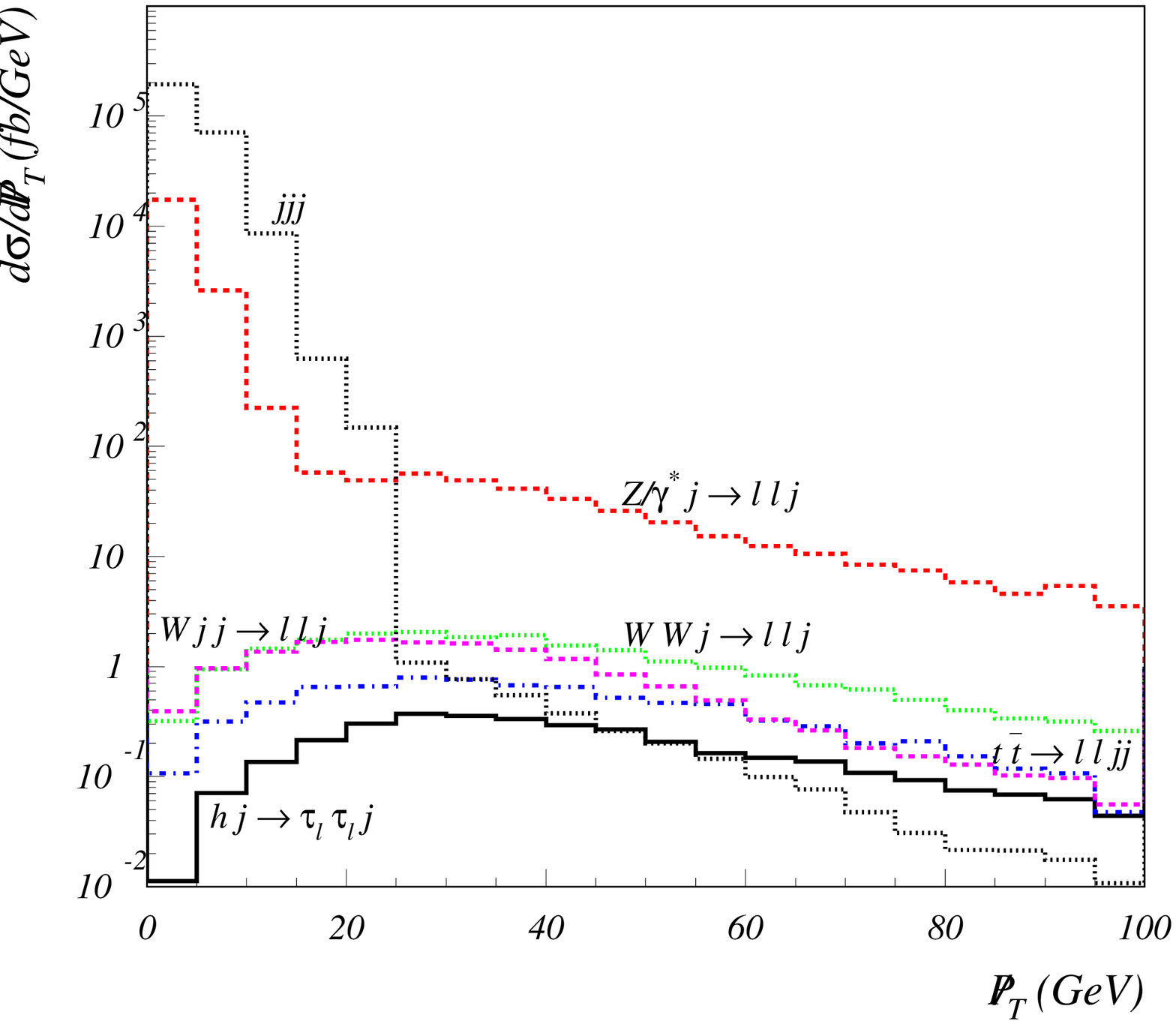}
\includegraphics[width=7.5cm]{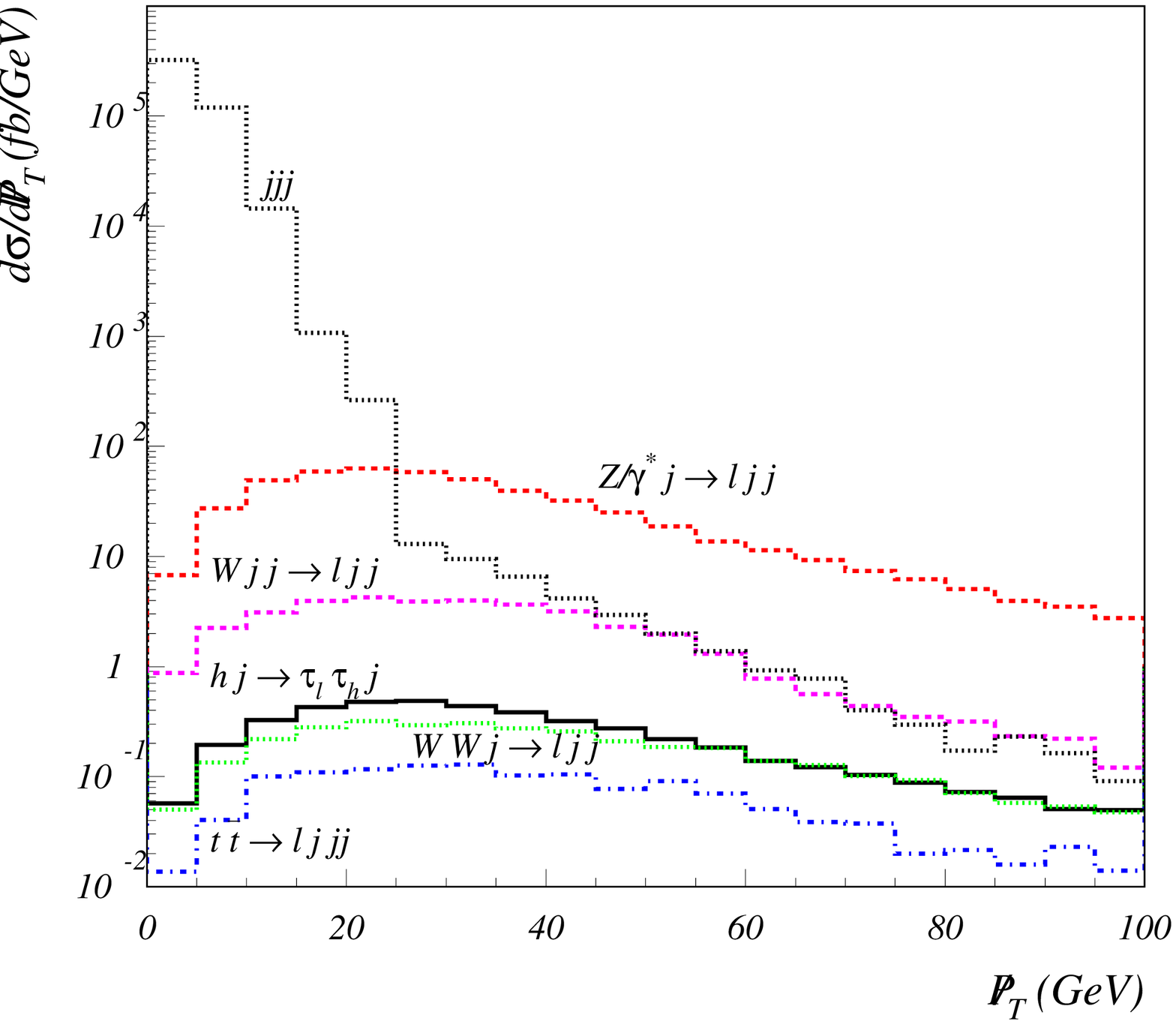}
\end{center}
%\vspace{-2cm}
\caption{Transverse missing energy distribution for signal and backgrounds for a Higgs mass of 120 $GeV$. On the left for the $ll$ analysis and on the right for the $lj$ case. In this figure all cuts described below were applied except for the missing energy cut.}
\label{fig:missing}
\end{figure}
%\end{widetext}
%%%%%%%%%%%%%%%%%%%%%%%%%%%%%%%%%%%%%%%%%%%%%%%%%%%%%%%%%%%%%%%%%%%%%%%%%%%%%%%%%%%

There are two other very important sources of reducible background. One is $pp \to e^+ e^- j (\mu^+ \mu^- j)$ in the case of the $ll$ final state and the other one is the huge QCD $pp \to jjj$ background with jets faking either $\tau$ final states or electron/muon final states. Fortunately, a judicial cut in the transverse missing energy will eliminate the first background. The QCD background has two distinct components: the one that does not contain heavy flavour jets and the one that does. The first one will again be eliminated by the transverse missing energy cut. The second one, due the semi-leptonic decays of the heavy quarks, has a significant increase on the total missing energy present. Therefore, a discussion on the probabilities of jet faking leptons and jets faking hadronic taus is in order.

Regarding the probabilities of jets faking either electrons or muons we will rely on MC studies performed by the CMS and ATLAS collaborations. CMS quotes a number of the order of $6 \times 10^{-4}$ for the probability of a jet faking an electron~\cite{CMS} while keeping the efficiency on electrons with $p_T$ from 5 to 50 GeV at the level of 90 \%. ATLAS~\cite{Aad:2009wy} quote numbers between $5 \times 10^{-4}$ and $10^{-5}$ for a central jet with $E_T > 17 $ GeV for electron reconstruction efficiencies between 77 \% and 64 \%. The numbers for the probability of a jet faking a muon are usually one order of magnitude smaller. However, the probabilities of b-jets faking electrons are usually higher, of the order of $10^{-3}$ and similar for electrons and muons. We have chosen a common conservative factor of $10^{-3}$ for the probability of a jet faking either an electron or a muon. As for the probability of a jet faking a hadronic tau, as explained earlier, we have taken the tau reconstruction efficiency to be 0.3 and accordingly we have used a tau rejection factor against jets as a function of the jet $p_T$~\cite{Aad:2009wy} that range from $10^{-2}$ to $10^{-3}$.

In Fig.~\ref{fig:missing} we present the transverse missing energy distribution, which is defined as the imbalance of all observed momenta (see discussion below), for a Higgs mass of 120 GeV. The three jet final state, including all possible combinations of light jets and b and c-jets, was generated with CalcHEP. All b and c-jets having three body decays are treated as follow: if one lepton and one hadron or two hadrons have $\Delta R < 0.4$ they are treated as one hadron; if the final state has only three particles our analysis applies, otherwise if there are four particles or more in the final state we discard those combinations where all jets have a $p_T$ above 20 GeV. The decays of b and c-quarks allow us to take into account the right amount of missing energy in the event - the b and c-jets contributions are then multiplied by the fake rates as described above. On the left plot, where the pure leptonic final state is shown, one can clearly see not only the $jjj$ background but also the $pp \to e^+ e^- j (\mu^+ \mu^- j)$ one. In the right plot we show the semi-leptonic scenario and only the $jjj$ background is important for low transverse missing energy. It is clear in both cases that, by requiring that $\slashed{E}_T > 30\, GeV$, both the $jjj$ and the $Z j \to e^+ e^- (\mu^+ \mu^-) j$ backgrounds are reduced to values well below the signal except for the tail due to the heavy flavour semi-leptonic decays.

In table~\ref{tab:iniend} we present the values for signal and background rates for a Higgs mass of 120 GeV. "Minimal Cuts" mean cuts which were applied to avoid soft and collinear divergences in the signal and in the background processes such a minimal $p_T$ cut of 20 $GeV$ on all jets (except for the $t \bar{t}$ one)
as well as jet separation $\Delta R_{jj} > 0.4$ for $Wjj$ and $jjj$ final states. We note that for the same set of cuts, the vector boson fusion process of Higgs production, for the same Higgs mass, is of the order 0.5 fb and therefore it can be safely neglected when compared to the signal (GGH) cross section. Again this is important because this way we are sure to be probing the Higgs coupling to fermions.

\begin{table}[ht]
\begin{center}
\begin{tabular}{c c c c c c c c c c c c c c c c} \hline \hline
\small{Process ($fb$)}  & $hj$ &  \small{$(Z/\gamma^* \to ll)  j$} & \small{$WWj$} & \small{$Wjj$} &  \small{$t \bar{t}$} &  \small{$jjj$} \\ \hline
\small{Minimal Cuts} & 1.2 $\times$ $10^3$ & 2.1 $\times$ $10^6$ & 8.7 $\times$ $10^4$  & 1.7 $\times$ $10^7$ & 8.3 $\times$ $10^5$ & 2.9 $\times$ $10^{10}$   \\ \hline
\small{Final ($ll$)}   &  13.8  &  93.8   &  13.0   &  17.2 & 4.7  & 2.9 \\ \hline
\small{Final ($lj$)}   &  14.1  &  83.9  &   2.3   &  56.8     & 0.7 & 23.7 \\ \hline \hline
\end{tabular}
\caption{Signal and each of the backgrounds in $fb$ for $m_h$ = 120 GeV. Signal and backgrounds in the minimal cut version were generated with $p_T^j > 20$ $GeV$ except for the $t \bar{t}$ background where no cut was applied. The $Z/\gamma^*$ background was generated with a cut in the leptons invariant mass of 5 GeV.}
\end{center}
\label{tab:iniend}
\end{table}

%%%%%%%%%%%%%%%%%%%%%%%%%%%%%%%%%%%%%%%%%%%%%%%%%%%%%%%%%%%%%%%%%%%%%%%%%%%%%%%%%%
%\begin{widetext}
\begin{figure}[h]
\begin{center}
\includegraphics[width=7.5cm]{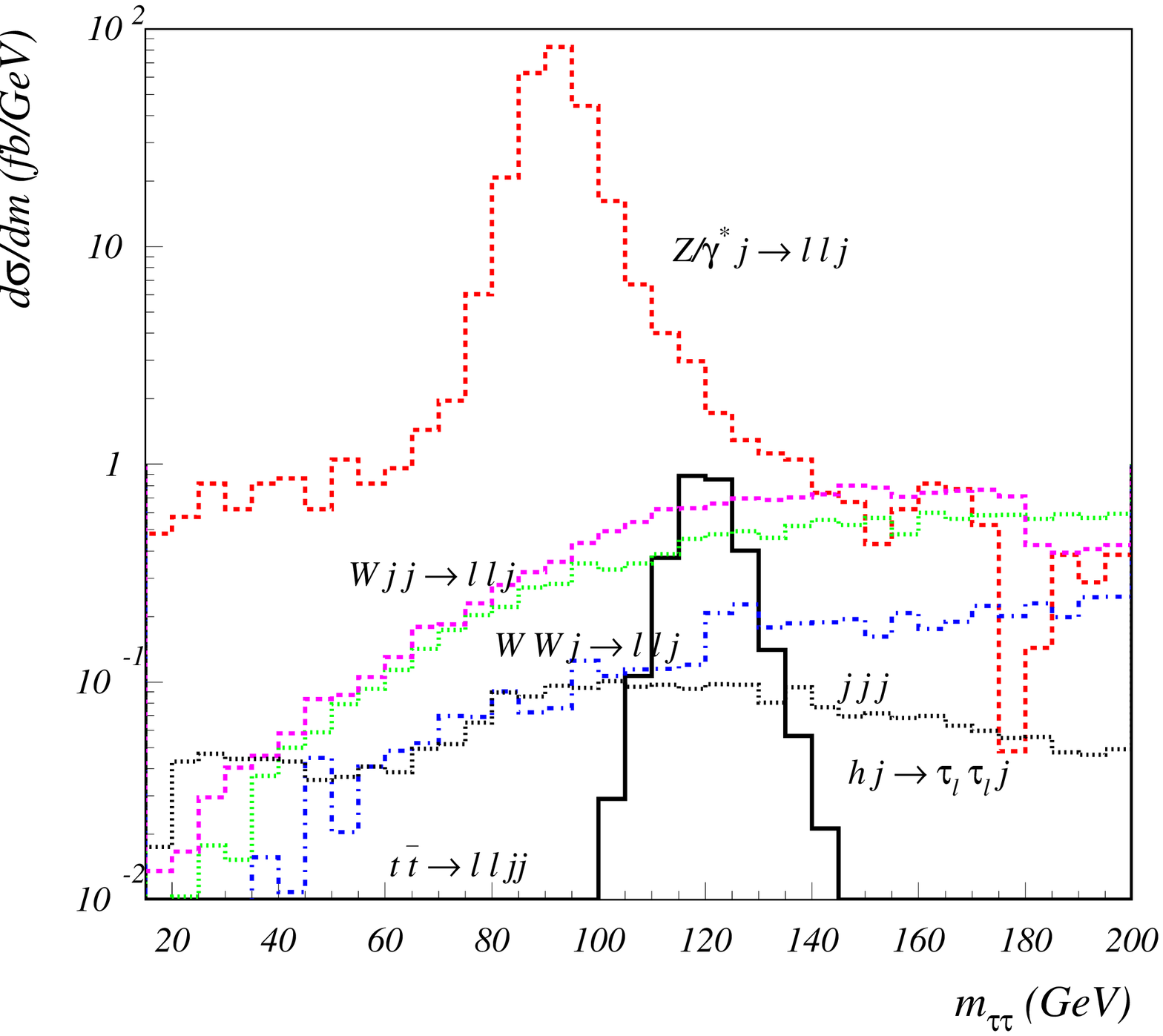}
\includegraphics[width=7.5cm]{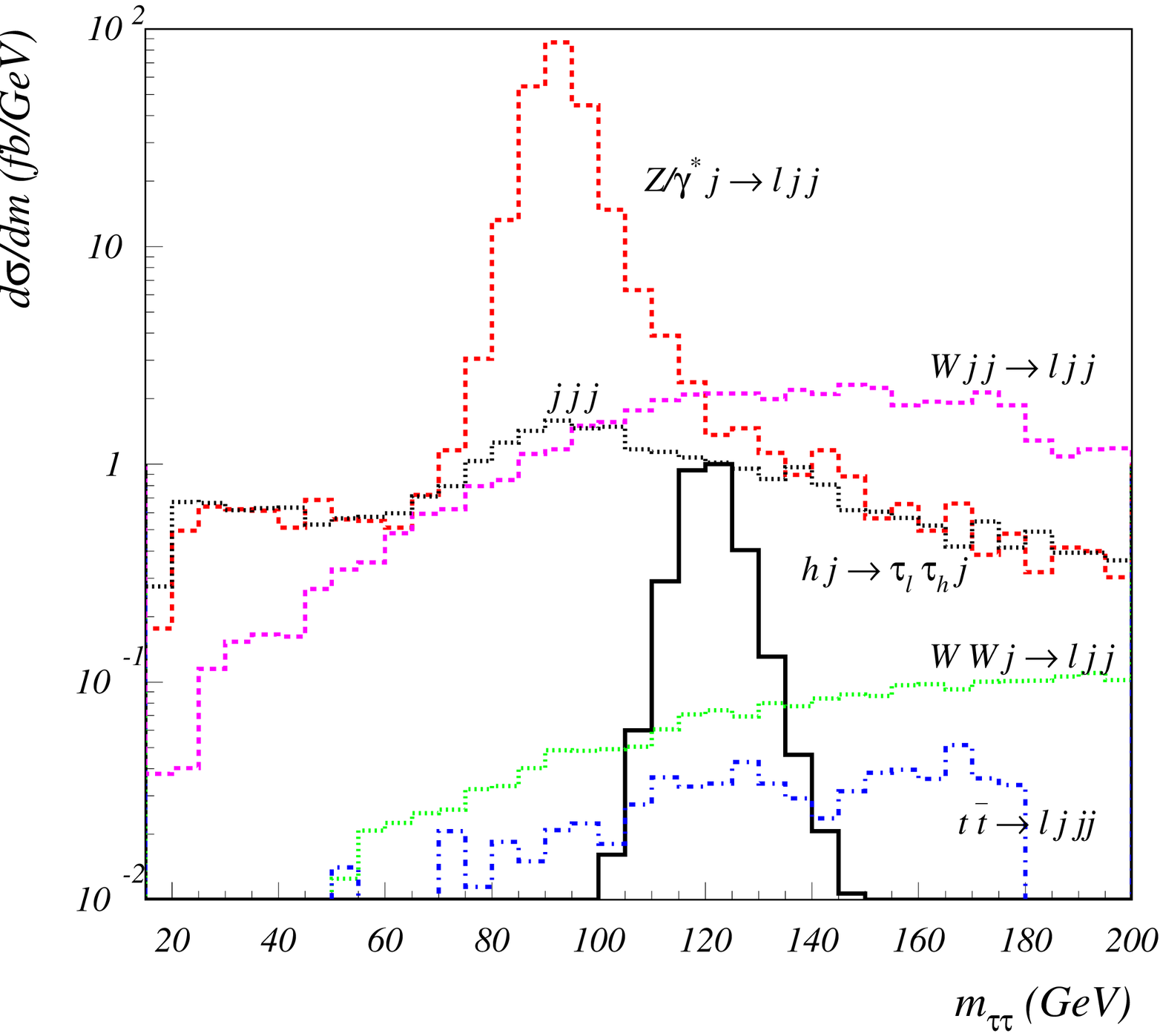}
\end{center}
%\vspace{-2cm}
\caption{Reconstructed mass $m_{\tau \tau}$ distributions for $\tau^+ \tau^-$ decaying leptonically  on the left and semi-leptonically on the right for a Higgs mass of 120 $GeV$. }

\label{fig:window}
\end{figure}
%\end{widetext}
%%%%%%%%%%%%%%%%%%%%%%%%%%%%%%%%%%%%%%%%%%%%%%%%%%%%%%%%%%%%%%%%%%%%%%%%%%%%%%%%%%%

A clear identification of the Higgs boson signal can only be accomplished with the reconstruction of the mass peak $m_{\tau \tau}$ at $m_h$ which would allow to effectively reduce the dominating $Zj$ background yielding a respective mass peak with  $m_{\tau \tau} \approx m_Z$. Due to the missing energy carried away by the two tau-neutrinos from tau-leptons decays, one can only effectively reconstruct the Higgs mass from the $\tau^+\tau^-$ decay products if the tau pairs are not back-to-back in the transverse plane~\cite{Ellis:1987xu}. As we are using a channel where the Higgs is produced alongside a high $p_T$ jet, the Higgs is also produced with a finite transverse momentum. Since $\tau^+ \tau^-$ pairs are ultra-relativistic, their decay products, including tau-neutrinos, are almost collinear with the $\tau$ direction. Therefore, the sizable momentum transfer of the Higgs boson in the process under consideration leads to an observable missing $E_T$ in the transverse plane. The measured $E_x$ and $E_y$ components allow us to reconstruct the missing momenta along the direction of each tau-lepton separately as done in~\cite{Belyaev:2002zz}, as a practical realisation of the ideas expressed in~\cite{Ellis:1987xu} for Higgs boson production at the Tevatron. This reconstruction enable us to finally form the Higgs mass peak from this decay to tau-leptons.

In Fig.~\ref{fig:window} we show the reconstructed $\tau^+ \tau^-$ invariant mass distribution for a Higgs mass of 120 $GeV$ after all cuts for the leptonic (left) and semi-leptonic (right) analysis. Signal and all backgrounds are identified in the figure. In both analyses we have sharp mass peaks for the signal. The reason is that what is lost in resolution in the semi-leptonic case because of the jet in the final state is somehow recovered by the fact that there is less missing energy involved and therefore the collinear approximation works better. Also clearly seen is the $Zj$ background peaking at $m_Z$, with a long tail after $m_Z$, that decreases dramatically as we move away from the peak and so will the total background until it stabilizes. All the above discussion can then be summarised and quantified in the following event selection
procedure.

\begin{itemize}
\item We require one electron with $p_T^e > 22 \, GeV$ or one muon with $p_T^\mu > 20 \, GeV$ for
triggering purposes. The additional lepton in the event has $p^{e}_T > 15 \, GeV$ and $p^{\mu}_T
> 10 \, GeV$. A 90 \% efficiency is assumed for the reconstruction of the electron and muon and
the separation between leptons and/or jets was chosen as $\Delta R_{j(l)j(l)} > 0.4$ and $|\eta_{l} | < 3.5$ for all leptons.
\item We require that at least one jet has $p^{j}_T > \,40 GeV$ and $|\eta_{j}| < 4$.
\item We require that the hadronic tau has $p^{j}_T > \,20 GeV$ and $|\eta_{j}| < 4$.
\item We veto the event if there is an additional jet with $p^{j}_T > \,20 GeV$ and $|\eta_{j}| < 5$.
\item We apply a mass window $m_h - 15 \, GeV < m_{\tau \tau} < m_h + 15 \, GeV$.
\item Events are vetoed if the tagging jet consistent with a $b$-jet hypothesis is found with $|\eta| < 2.5$
(we assume a $b$-jet tagging efficiency of 60 \%).
\item Finally we require the transverse missing energy to be $\slashed{E}_T > 30 \, GeV$.
\end{itemize}

In table~\ref{tab:sigbac} we present the signal and sum of all background cross sections, signal-to-background ($\sigma_{S}/\sigma_{B} $) ratios and $\sigma_{S}/\sqrt{\sigma_{B}} $ as a function of the Higgs mass. In the first and second columns we show the results for the $ll$ analysis while columns three and four are for the $lj$ case. In the last two columns we present the combined values, summed under quadrature, for the signal-to-background $\sigma_{S}/\sigma_{B} $ ratio and the sensitivity $\sigma_{S}/\sqrt{\sigma_{B}} $. The analysis was done for mass values as low as 100 $GeV$ so that they can be used to explore extensions of the SM where such a light Higgs is not excluded yet. The values for masses above 150 $GeV$ are also shown because they can be very important for models where the decays to gauge bosons are suppressed.

It is clear that the signal observation can be systematically challenging but the values of the $\sigma_{S}/\sigma_{B} $ ratio can be improved at the expense of the significance by shrinking the Higgs mass window especially when its mass is close to the mass of the $Z$ boson. The highest significance and the highest $\sigma_{S}/\sigma_{B} $ ratio takes place at $m_h=120$ GeV and $m_h=130$ $GeV$ respectively, because it is sufficiently distant from the irreducible $Z j $ background peak while the effect of the decay to gauge bosons does not fully come into play  yet. In table~\ref{tab:lumi} we present the luminosities required for a 95 \% CL exclusion, 3$\sigma$ and 5$\sigma$ discovery of a SM Higgs boson at $\sqrt{s} = 14$  TeV as a function of the Higgs mass. A 95 \% exclusion limit for the SM Higgs boson in the mass range 120--140 $GeV$ would require less than 3 $fb^{-1}$ of total integrated luminosity.

\begin{table}[ht]
\begin{center}
\begin{tabular}{c c c c c c c c c c c c c c c c c c c c } \hline \hline
 $m_h$ \small{($GeV$)} & $\sigma_{S_{(ll)}}$ \small{($fb$)}   & $\sigma_{B_{(ll)}}$ \small{($fb$)}&  $\sigma_{S_{(lj)}}$ \small{($fb$)}   & $\sigma_{B_{(lj)}}$ \small{($fb$)}& $\sigma_{S}/\sigma_{B}$ \small{(\%)} & $\sigma_{S}/\sqrt{\sigma_{B}}$  \small{($\sqrt{fb}$ )}  \\
\hline
100        &  14.9      &    1162.6   &  15.5      &    1156.3  & 1.85    &  0.63  \\ \hline
110        &  15.1      &    437.4    &  15.4      &    451.0   & 4.86    &  1.02  \\ \hline
120        &  13.8      &    131.7    &  14.1      &    167.4   & 13.5    &  1.62  \\ \hline
130        &  10.4      &    90.9     &  11.5      &    129.7   & 14.5    &  1.49  \\ \hline
140        &  6.7       &    78.0     &  7.5       &    117.7   & 10.7    &  1.03  \\ \hline
150        &  3.3       &    71.3     &  3.8       &    107.7   & 5.89    &  0.54  \\ \hline
160        &  0.74      &    66.5     &  0.87      &    98.4    & 1.42    &  0.13  \\ \hline
170        &  0.16      &    60.6     &  0.19      &    90.4    & 0.34    &  0.029  \\ \hline
180        &  0.098     &    59.6     &  0.12      &    86.4    & 0.22    &  0.018  \\ \hline
190        &  0.060     &    58.8     &  0.075     &    80.7    & 0.14    &  0.011  \\ \hline
200        &  0.044     &    58.7     &  0.055     &    76.9    & 0.10    &  0.0085  \\ \hline \hline
\end{tabular}
\caption{Cross sections for signal and sum of all backgrounds after all cuts as a function of the Higgs mass for a SM Higgs boson. In the first and second columns we show the results for the $ll$ analysis while columns three and four are for the $lj$ case. In the last two columns we present the combined values for $\sigma_{S}/\sigma_{B}$ and $\sigma_{S}/\sqrt{\sigma_{B}}$, summed under quadrature. The analysis was done for mass values as low as 100 $GeV$ so that they can be used to explore extensions of the SM where such a light Higgs state is not excluded yet.}
\end{center}
\label{tab:sigbac}
\end{table}
\begin{table}[ht]
\begin{center}
\begin{tabular}{c c c c c c c c c c c c c c c c c c c} \hline \hline
$m_h$ \small{($GeV$)} & \small{95 \% CL exclusion  $L (fb^{-1})$} & \small{3$\sigma$ discovery $L (fb^{-1})$} &  \small{5$\sigma$ discovery $L (fb^{-1})$} \\
\hline
100        &  10      &    23   &   63   \\ \hline
110        &  3.8     &    8.6  &   24   \\ \hline
120        &  1.5     &    3.4  &   9.5  \\ \hline
130        &  1.8     &    4.1  &   11   \\ \hline
140        &  3.8     &    8.6  &   24   \\ \hline
150        &  14      &    31   &   85   \\ \hline \hline
\end{tabular}
\caption{Integrated luminosities needed to reach a 95 \% CL exclusion,
3$\sigma$ and 5$\sigma$ discovery for a SM Higgs boson at the LHC. Luminosities
shown are for the combined results of the two analysis.}
\end{center}
\label{tab:lumi}
\end{table}

\section{The Leptonic 2HDM}
\label{sec:2HDM}

In this section we give a brief description of the leptonic 2HDM and some of its extensions. We consider the eight parameter 2HDM Higgs potential which is invariant under the $Z_2$ discrete symmetry $\Phi_1 \to \Phi_1$, $\Phi_2 \to - \Phi_2$, softly broken by the term $[m_{12}^2\Phi_1^\dagger\Phi_2+{\rm h.c.}]$. The vacuum structure is chosen such that the potential does not break either CP or the electric charge spontaneously and the potential is written as
\begin{eqnarray}
V(\Phi_1,\Phi_2) &=& m^2_1 \Phi^{\dagger}_1\Phi_1+m^2_2
\Phi^{\dagger}_2\Phi_2 - (m^2_{12} \Phi^{\dagger}_1\Phi_2+{\rm
h.c}) +\frac{1}{2} \lam_1 (\Phi^{\dagger}_1\Phi_1)^2 +\frac{1}{2}
\lam_2 (\Phi^{\dagger}_2\Phi_2)^2\nonumber \\ &+& \lam_3
(\Phi^{\dagger}_1\Phi_1)(\Phi^{\dagger}_2\Phi_2) + \lam_4
(\Phi^{\dagger}_1\Phi_2)(\Phi^{\dagger}_2\Phi_1) + \frac{1}{2}
\lam_5[(\Phi^{\dagger}_1\Phi_2)^2+{\rm h.c.}] ~, \label{higgspot}
\end{eqnarray}
where $\Phi_i$, $i=1,2$, are complex $SU(2)$ doublets with Vacuum Expectation Values (VEVs) $<\Phi_i> = v_i$ and all parameters are real. The model has eight degrees of freedom which after spontaneous symmetry breaking give rise to the three Goldstone bosons partners of the weak gauge bosons and five Higgs states, two CP-even, $h$ and $H$, one CP-odd, $A$ and two charged Higgs boson, $H^{\pm}$. As one parameter is fixed by the electroweak breaking scale, there are still seven independent parameters and we adopt the following set: the four masses $m_{h}$, $m_{H}$, $m_{A}$ and $m_{H^\pm}$, the two angles $\tan\beta = v_2/v_1$ and $\alpha$ and $M^2= m_{12}^2/(\sin \beta \cos \beta)$, which is a measure of the discrete symmetry breaking. The angle $\beta$ is the rotation angle from the group eigenstates to the mass eigenstates in the CP-odd and charged sectors. The angle $\alpha$ is the corresponding rotation angle for the CP-even sector.

The most general Yukawa Lagrangian that one can build in a 2HDM originates Flavour Changing Neutral Currents (FCNCs) at tree-level. The simplest way to avoid FCNCs is to extend the $Z_2$ symmetry to the fermions. It suffices that fermions of a given electric charge couple to no more than one Higgs doublet~\cite{Glashow}. This can be accomplished naturally by imposing on all fields appropriate discrete symmetries that forbid the unwanted FCNC couplings. There are essentially four possible independent combinations~\cite{catalogue}: type I is the model where only the doublet $\phi_2$ couples to all fermions; type II is the model where $\phi_2$ couples to up-type quarks and $\phi_1$ couples to down-type quarks and leptons; in a Type III model $\phi_2$ couples to up-type quarks and to leptons and $\phi_1$ couples to down-type quarks; a Type IV model is instead built such that $\phi_2$ couples to all quarks and $\phi_1$ couples to all leptons. This last model is also called the leptonic 2HDM and was first discussed in~\cite{L2HDM1}. A study on the model's phenomenology at the LHC was recently presented in~\cite{Su:2009fz}. The couplings of the lightest CP-even Higgs to the fermions in this model are
\begin{equation}
\overline{l} l h: \quad \frac{ig}{2 m_W} \frac{\sin
\alpha}{\cos \beta} m_{l}
\qquad \qquad \overline{q} q h:  \quad -\frac{ig}{2 m_W}\frac{\cos
\alpha}{\sin \beta} m_{q}
\end{equation}
where $q$ stands for a quark and $l$ stands for a lepton. For the heavier CP-even Higgs, $H$, one has
\begin{equation}
\overline{l} l H: \quad - \frac{ig}{2 m_W} \frac{\cos
\alpha}{\cos \beta} m_{l}
\qquad \qquad \overline{q} q H:  \quad -\frac{ig}{2 m_W}\frac{\sin
\alpha}{\sin \beta} m_{q}
\end{equation}
while the couplings of $h$ and $H$ to the gauge bosons relative to the SM Higgs couplings are universal and given by
\begin{equation}
V V h \quad g_{_{VVh_{SM}}} \sin (\beta -\alpha)
\qquad \qquad
V V H  \quad g_{_{VVh_{SM}}} \cos (\beta -\alpha) \enskip .
\end{equation}

The leptonic 2HDM has a phenomenological advantage relative to the more popular model II - the results obtained for the SM can be used for the leptonic 2HDM with little changes. When the decay to gauge bosons is kinematically forbidden and simultaneously both CP-even Higgses are not allowed to decay to final states where another lighter Higgs is present, both CP-even Higgs states decay mainly to $b \bar{b}$ and $\tau^+ \tau^-$. The decay to $c \bar{c}$ and to gluon pairs can also play an important role and each one of these decays can even be the most important signature in certain regions of the parameter space~\cite{Arhrib:2009hc}. In the leptonic 2HDM, the Higgs couples to up and down quarks with the same strength. Hence, there is a common factor that factorises for quarks and another factor that factorises for leptons. If we neglect decays to very light leptons and to a photon pair we can write the BRs of the two CP-even Higgs states as a function of the SM Higgs
BRs to $\tau^+ \tau^-$ as
\begin{equation}
BR(h \to \tau^+ \tau^-) \simeq \frac{\frac{\sin^2 \alpha}{\cos^2 \beta} \, {\rm{BR}}( h_{SM} \to \tau^+ \tau^-)}{(\frac{\sin^2 \alpha}{\cos^2 \beta}-\frac{\cos^2 \alpha}{\sin^2 \beta}) \, {\rm{BR}}( h_{SM} \to \tau^+ \tau^-) + \frac{\cos^2 \alpha}{\sin^2 \beta}}
\label{eq:Br1}
\end{equation}
and
\begin{equation}
BR(H \to \tau^+ \tau^-)  \simeq \frac{\frac{\cos^2 \alpha}{\cos^2 \beta} \, {\rm{BR}}( h_{SM} \to \tau^+ \tau^-)}{(\frac{\cos^2 \alpha}{\cos^2 \beta}-\frac{\sin^2 \alpha}{\sin^2 \beta}) \, {\rm{BR}}( h_{SM} \to \tau^+ \tau^-) + \frac{\sin^2 \alpha}{\sin^2 \beta}} \enskip .
\label{eq:Br2}
\end{equation}
The same reasoning applies to the cross sections. For VBF we just have to multiply by $\sin^2 (\beta - \alpha)$ in eq.~(\ref{eq:Br1}) and $\cos^2 (\beta - \alpha)$ in eq.~(\ref{eq:Br2}) to get $\sigma(qq' \to h (H) qq') \, {\rm{BR}}(h (H) \to \tau^+ \tau^-)$ as a function of the corresponding SM cross section and SM Higgs BR to $\tau^+ \tau^-$. For the production process $pp \to gg \to h j \to h \tau^+ \tau^-$ the relation between the leptonic 2HDM and the SM can explicitly be written as
\begin{equation}
\sigma(gg \to h g) \, {\rm{BR}}(h \to \tau^+ \tau^-) \simeq \sigma(gg \to h_{SM} g) \, \frac{\frac{\sin^2 \alpha}{\cos^2 \beta} \, {\rm{BR}}(h_{SM} \to \tau^+ \tau^-)}{BR(h_{SM} \to \tau^+ \tau^-) (\tan^2 \beta \tan^2 \alpha -1)+1}
\end{equation}
and
\begin{equation}
\sigma(gg \to Hg) \, {\rm{BR}}(H \to \tau^+ \tau^-) \simeq \sigma(gg \to h_{SM} g) \, \frac{\frac{\cos^2 \alpha}{\cos^2 \beta} \, {\rm{BR}}(h_{SM} \to \tau^+ \tau^-)}{BR(h_{SM} \to \tau^+ \tau^-) (\frac{\tan^2 \beta}{\tan^2 \alpha} -1)+1} \quad .
\end{equation}
These expressions tell us how to translate the results obtained for the SM to the leptonic 2HDM. In the next sections we will apply the results to different models introduced recently and that have the leptonic 2HDM as a submodel. We will further show that such leptonic 2HDMs can be probed in the low luminosity regime.
\subsection{The AKS Model}
\label{sec:AKS}

Recently, Aoki, Kanemura and Seto (AKS)~\cite{Aoki:2008av, Aoki:2009vf} have proposed a model to account for neutrino oscillation, DM and baryon asymmetry of the Universe at the TeV scale and without introducing very high mass scales. The scalar sector of the model is a leptonic 2HDM plus a charged singlet (a Zee model~\cite{Zee:1980ai}) plus a neutral singlet. Besides the usual 2HDM $Z_2$ symmetry, that when extended to the fermions prevents tree-level FCNCs, a new $Z_2$ symmetry is introduced under which the new particles arising from the singlets have odd parity whereas gauge and matter fields and the 2HDM fields are even. Tiny neutrino masses are generated at the three loop level due to the extra $Z_2$ symmetry which also guarantees the stability of a DM candidate. A discussion on the collider phenomenology of the model can be found in~\cite{Aoki:2009ha}.
%
%\begin{widetext}
\begin{figure}
\begin{center}
\includegraphics[width=7.8cm]{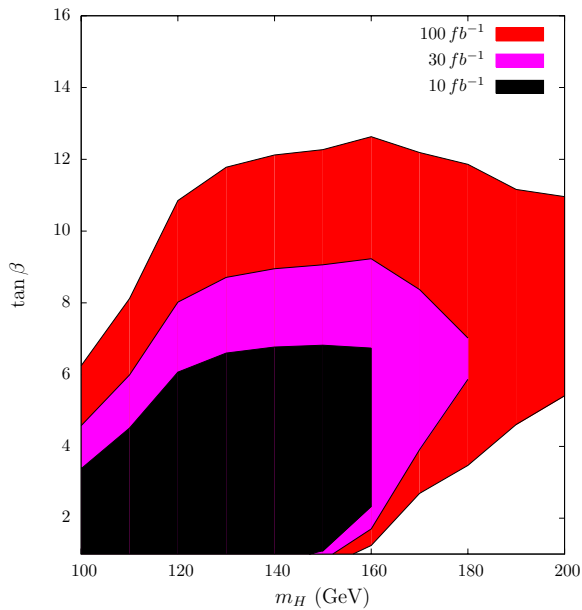}
\end{center}
%\vspace{-2cm}
\caption{95 \% CL sensitivity regions for the AKS model in the ($\tan \beta$, $m_H$) plane for three different luminosities. $m_H$ is the non-standard model like Higgs boson.}
\label{fig:AKS}
\end{figure}
%\end{widetext}
%

There are particular scenarios where the parameters of the model are such that new physics can be accommodated without fine tuning. The LEP bound forces $m_{H^\pm} \ga 90$ \,$GeV$ and the global custodial symmetry is exact if $m_{H^\pm} \approx m_H$ and $\sin (\beta - \alpha) \approx 1$ (we will discuss all experimental and theoretical conditions in the next section). Combining this with the requirement of natural generation of tiny neutrino masses they obtain $m_{H^\pm} = m_{H} \approx 100$ $GeV$. With these constraints these are perfect candidates to search for at the LHC. As the decay to leptons is enhanced, all processes with leptons in the final state are the ones that will give a more clear signature of the model. In the scenario discussed, $h$ has SM-like couplings to fermions and gauge bosons. Hence, the SM results can be used to find or constrain such a Higgs state. The other CP-even Higgs state, $H$, has no couplings to the gauge bosons because $\sin (\beta - \alpha) \approx 1$. Therefore the decay $H \to \tau^+ \tau^-$ is now dominant for all the range of Higgs mass considered  - our analysis extends to a Higgs mass of 200 $GeV$ although in this particular model masses $O (100)$ GeV are preferred. It is important to note that this is exactly the case where the complementarity discussed in the introduction shows. Although similar in some senses, the VBF and gluon fusion processes give very different results in this scenario: VBF is close to zero in its yield, while gluon fusion gives a number of events well above the SM scenario.

The condition $\sin (\beta - \alpha) \approx 1 $ fixes $\alpha$ and therefore we will present our results as a function of $\tan \beta$. It is possible to write the BR for the new scalar $H$ as a function of the BR of the SM Higgs decaying in $\tau^+ \tau^-$,
\begin{equation}
{\rm{BR}}(H \to \tau^+ \tau^-) \simeq \frac{\tan^2 \beta \, \Gamma_{h_{SM} \to \tau^+ \tau^-}}{\tan^2 \beta \, \Gamma_{h_{SM} \to \tau^+ \tau^-} + \cot^2 \beta \,(\Gamma_{h_{SM}} -\Gamma_{h_{SM} \to \tau^+ \tau^-})}
\end{equation}
where the equality is not exact because there are small contributions from the photon width and from the muon pair channel. This allows us to write
\begin{equation}
\sigma(gg \to Hg) \, {\rm{BR}}(H \to \tau^+ \tau^-) \simeq \sigma(gg \to h_{SM} g) \, \frac{\tan^2 \beta \, {\rm{BR}}(h_{SM} \to \tau^+ \tau^-)}{BR(h_{SM} \to \tau^+ \tau^-) (\tan^4 \beta -1)+1}.
\end{equation}

In Fig.~\ref{fig:AKS} we present the region in the ($\tan \beta$, $m_H$) plane that can be probed at the LHC at 95 \% CL for three different sets of luminosities. There is a strong dependence on the Higgs boson mass but nevertheless a considerable range of $\tan \beta$ is still accessible. As discussed in~\cite{Aoki:2009vf}, small $\tan \beta$ ($1 \lesssim \tan \beta \lesssim 10$) is favoured by EW baryogenesis. The region of $\tan \beta$ that can be probed at the LHC lies in the favoured domain but even with $100 fb^{-1}$ there will still be a region of small masses that will not be covered. This region will most probably be covered by searches for a light charged Higgs state with mass around 100 $GeV$. This process is better designed to cover the mass regions above 120--130 $GeV$.

\subsection{Limiting Scenarios of the Leptonic 2HDM}
\label{sec:limit}

%
%\begin{widetext}
\begin{figure}
\begin{center}
\includegraphics[width=7.5cm]{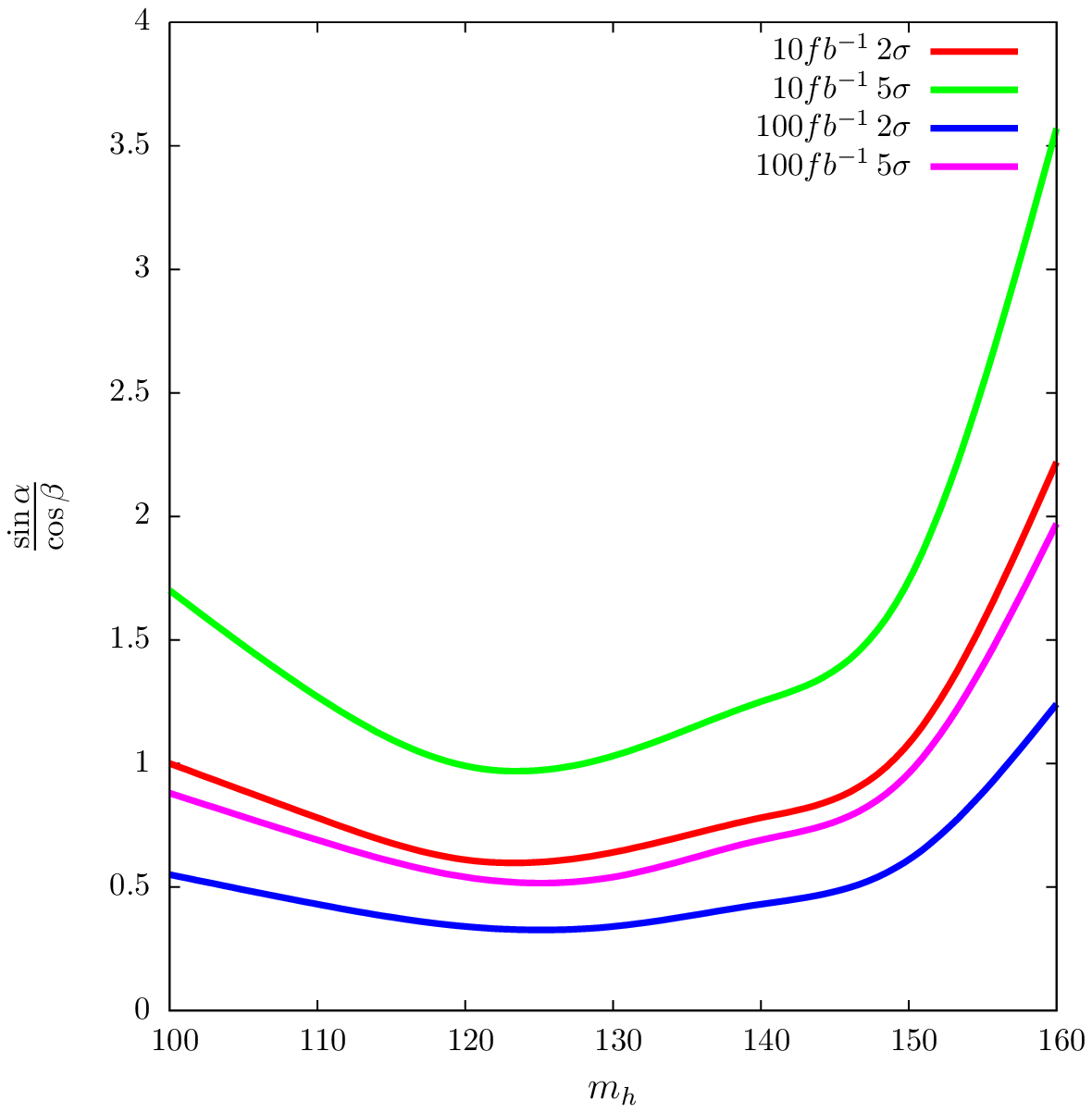}
\includegraphics[width=7.5cm]{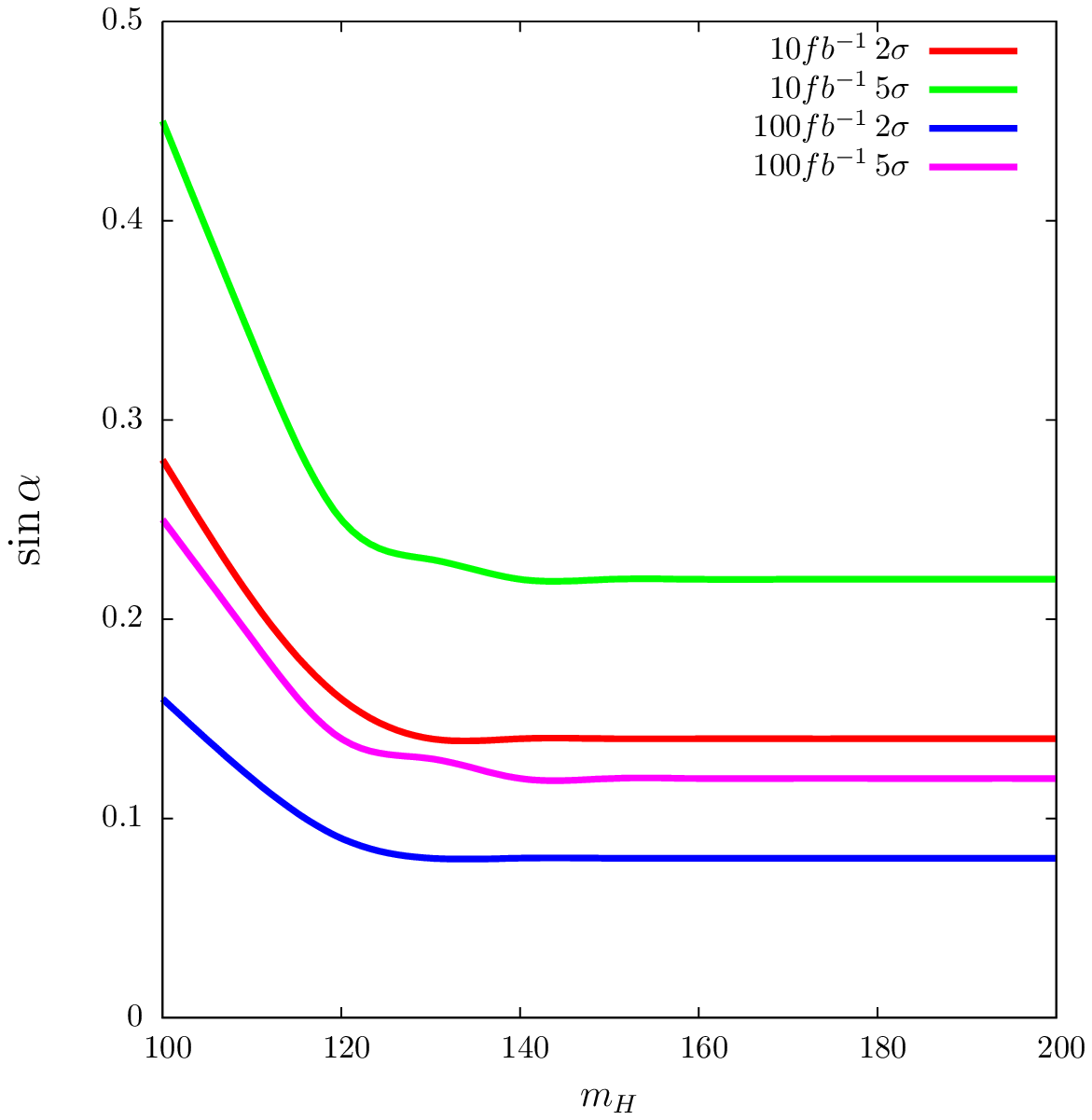}
\end{center}
%\vspace{-2cm}
\caption{In the left panel we show the signal significance contours for the models in~\cite{Goh:2009wg} in the ($\sin \alpha/\cos \beta$, $m_h$) plane for $2\sigma$ and $5\sigma$ and for two different sets of luminosities. In the right panel we show the significance contours in the ($\sin \alpha$, $m_H$) plane.}
\label{fig:Hall}
\end{figure}
%\end{widetext}
%

Two extensions of the leptonic 2HDM were recently discussed in~\cite{Goh:2009wg}: the singlet DM model and the inert Higgs doublet model (IDM). The first is a leptonic 2HDM plus one neutral singlet while the second is a three doublet model where one of the doublets does not acquire a vacuum expectation value. The singlet in the first model and the inert doublet in the second model provide dark matter candidates. In the scenario discussed in~\cite{Goh:2009wg}, the doublet that gives mass to the leptons has a vacuum expectation value much smaller than the one that gives mass to the quarks, which in our notation can be translated into
\begin{equation}
v_1 \ll v_2 \quad \Rightarrow \quad \tan \beta \gg 1 \enskip ,
\label{eq:L2HDM1}
\end{equation}
and if additionally we force one of the CP-even Higgs, $H$, to lie dominantly in the leptonic Higgs doublet, primary cosmic rays will be suppressed.
% $\bar{p}$.
 The latter condition reads
\begin{equation}
\sin \alpha \ll 1 \quad \Rightarrow \quad \tan \alpha \ll 1 \enskip .
\label{eq:L2HDM2}
\end{equation}
The other CP-even Higgs, $h$, is SM-like except in its couplings to the leptons. See~\cite{Goh:2009wg} for a detailed discussion. Both eqs.~(\ref{eq:L2HDM1}) and (\ref{eq:L2HDM2}) are satisfied in the limit $\beta - \alpha \approx \pi/2$ which is equivalent to the AKS scenario discussed in the previous section. Another possibility discussed in~\cite{Goh:2009wg} to enhance a leptonic signal is to keep $\sin \alpha$ and $\cos \beta$ small while allowing a large $\sin \alpha/\cos \beta$ ratio, that
is, $\sin \alpha \gg \cos \beta$ and $\cos \alpha \approx \sin \beta \approx 1$. Interestingly, in this limit, VBF~\cite{Goh:2009wg} and gluon fusion have the same limiting expressions as a function of the respective SM cross sections, that can be written as
\begin{equation}
\sigma(pp \to h g) \, {\rm{BR}}(h \to \tau^+ \tau^-) \simeq \sigma(pp \to h_{SM} g) \, \frac{\frac{\sin^2 \alpha}{\cos^2 \beta} \, {\rm{BR}}(h_{SM} \to \tau^+ \tau^-)}{BR(h_{SM} \to \tau^+ \tau^-) (\frac{\sin^2 \alpha}{\cos^2 \beta} -1)+1} \, ,
\end{equation}
and because in this limit BR$(H \to \tau^+ \tau^-) \approx 1$ one has
\begin{equation}
\sigma(pp \to Hg) \, {\rm{BR}}(H \to \tau^+ \tau^-) \simeq \sigma(pp \to h_{SM} g) \, \sin^2  \alpha \, ,
\end{equation}
where we have also used $\sin^2  (\alpha-\beta) \approx \sin^2  \alpha$. Note that the plots can be shown as a function of any of the two variables $\sin  (\alpha-\beta)$ or $\sin  \alpha$.

In the left panel of Fig.~\ref{fig:Hall} signal significance contours for the leptonic 2HDM scenarios discussed in~\cite{Goh:2009wg} are shown in the ($\sin \alpha/\cos \beta$, $m_h$) plane for 2$\sigma$ and 5$\sigma$ and for two different sets of luminosities. In the right panel we show the significance contours in the ($\sin \alpha$, $m_H$) plane. In \cite{Goh:2009wg} similar plots were shown based on the analysis for VBF presented in~\cite{Aad:2009wy} for $30 fb^{-1}$. Our significance curves are very close to the ones presented in \cite{Goh:2009wg} for the SM-like Higgs boson. In the ($\sin \alpha$, $m_H$) plane our results are not only slightly better but we have also extended them to a wider region of masses. The prospect of finding a signal for this particular model seems very promising.

\subsection{The general leptonic 2HDM at low luminosity}

As discussed in the introduction, there are four different Yukawa versions of a 2HDM without tree-level FCNCs. In models I and II the BR$(h \to \tau^+ \tau^-)$ is either the same or smaller than the corresponding SM BR. Therefore, the results are the ones discussed for the SM in most of the parameter space except in model II close to $\sin \alpha \approx 0$ when BR$(h \to \tau^+ \tau^-) \approx 0$. In model III there is always a region in the vicinity of $\sin \alpha \approx 0$, even if small, where BR$(h \to \tau^+ \tau^-)$ is dominant. For $\tan \beta = 1$ and a Higgs mass of 120 $GeV$ one has
$\sigma ({pp \to hg}) \, {\rm{BR}}({h \to \tau^+ \tau^-})/ (\sigma({pp \to h_{SM} g}) \, {\rm{BR}}({h_{SM} \to \tau^+ \tau^-})) \approx 10$. As $\tan \beta$ grows, the previous ratio becomes $\approx 3$ independently of the value of $\tan \beta$ while the region around $\sin \alpha = 0$ where the ratio is important shrinks. Therefore, all values of $\tan \beta$ will be probed in this version of the model, for a light Higgs and $|\sin \alpha| \lesssim 0.2$.

\begin{figure}
%\begin{center}
\includegraphics[width=10.0cm]{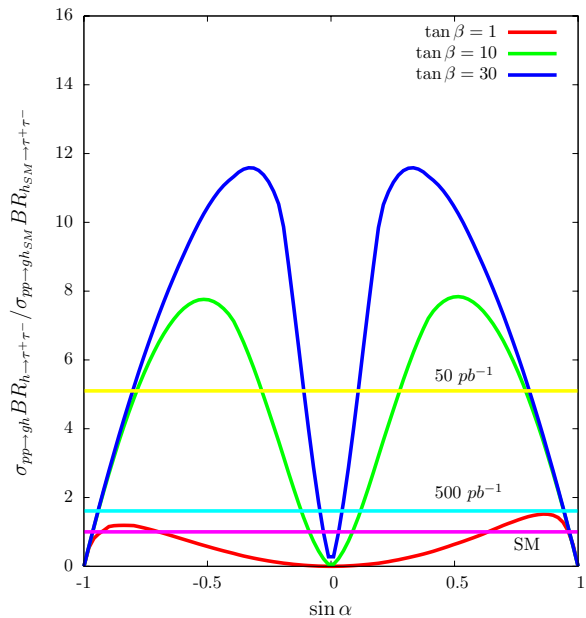}
%\includegraphics[width=7.6cm]{Hallpaper.eps}
%\end{center}
%\vspace{-2cm}
\caption{Ratio of leptonic 2HDM $\sigma({pp \to hg}) \, {\rm{BR}}({h \to \tau^+ \tau^-})$ to the SM $\sigma({pp \to h_{SM} g}) \, {\rm{BR}}({h_{SM} \to \tau^+ \tau^-})$ for $m_h = 120 \, GeV$ and three values of $\tan \beta$. Also shown are the 95 \% CL sensitivity lines for 50 $pb^{-1}$ and 500 $pb^{-1}$.}
\label{fig:modIV}
\end{figure}

In Fig.~\ref{fig:modIV}, $\sigma({pp \to hg}) \, {\rm{BR}}({h \to \tau^+ \tau^-})/ (\sigma({pp \to h_{SM} g}) \, {\rm{BR}}({h_{SM} \to \tau^+ \tau^-}))$ is presented for the leptonic model and a Higgs mass of 120 $GeV$. We have considered three values for $\tan \beta$ (1, 10, 30) and then varied $\sin \alpha$ between $-1$ and $1$. We have drawn 95 \% CL sensitivity lines for two luminosities (50 $pb^{-1}$ and 500 $pb^{-1}$). It is very important to note that these results do not depend on any of the 2HDM parameters besides $\sin \alpha$, $\tan \beta$ and the Higgs mass. The only assumption made is that the light Higgs is not allowed to decay to other final states with Higgs bosons. If, for example, $h \to AA$ is possible, the overall picture can change dramatically. It is clear from Fig.~\ref{fig:modIV} that even with only $\approx \, 50 \, pb^{-1}$ the large values of $\tan \beta$ are accessible for a considerable range of $\sin \alpha$.

%
%\begin{widetext}
%\end{widetext}

\section{Experimental and theoretical constraints}
\label{sec:bounds}

Except for the SM, all models discussed in this work have the leptonic 2HDM as a submodel. Most of the experimental bounds are therefore directly applicable to these more general models. Regarding the theoretical constraints a more detailed analysis is needed. In this work, all couplings relevant for the production and decay of the CP-even Higgs boson originate from the Yukawa Lagrangian\footnote{There are marginal contributions related to the structure of the Higgs potential that enter the total Higgs width via loops like for instance in $h \to \gamma \gamma$.}. The two most restrictive theoretical constraints are the ones arising from demanding vacuum stability and tree-level unitarity of the potential. First, the potential has to be bounded from below at tree-level. In a general 2HDM this is attained by imposing the constraints described in~\cite{vac1} to the parameters of the potential. A complete general description for all extensions of the 2HDM is not available but there is a detailed study of the case of a 2HDM plus one inert singlet~\cite{Grzadkowski:2009bt}. Bounds from perturbative unitarity are also available only for the 2HDM~\cite{unitarity}. For the parameter region we are probing, this set of constraints is enough for our study. We will also force perturbativity on the parameters of the potential by choosing $|\lambda_i|<8 \pi$. The requirement that the $\tau$ Yukawa coupling remains perturbative gives no useful bound on $\tan \beta$~\cite{Logan:2009uf} contrary to model type II where a similar requirement on the top and bottom Yukawa couplings forces $0.3\leq \tan\beta \leq 100$~\cite{catalogue}. The 2HDM vacuum is naturally protected against charge and CP breaking~\cite{charge}. The Zee model and consequently the AKS model is no longer protected. However, it was shown in~\cite{Barroso:2005hc} that no useful bounds could be derived for the parameters of the potential by forcing the vacuum structure of the potential to conserve electric charge.

\begin{figure}
\begin{center}
\includegraphics[width=7.4cm]{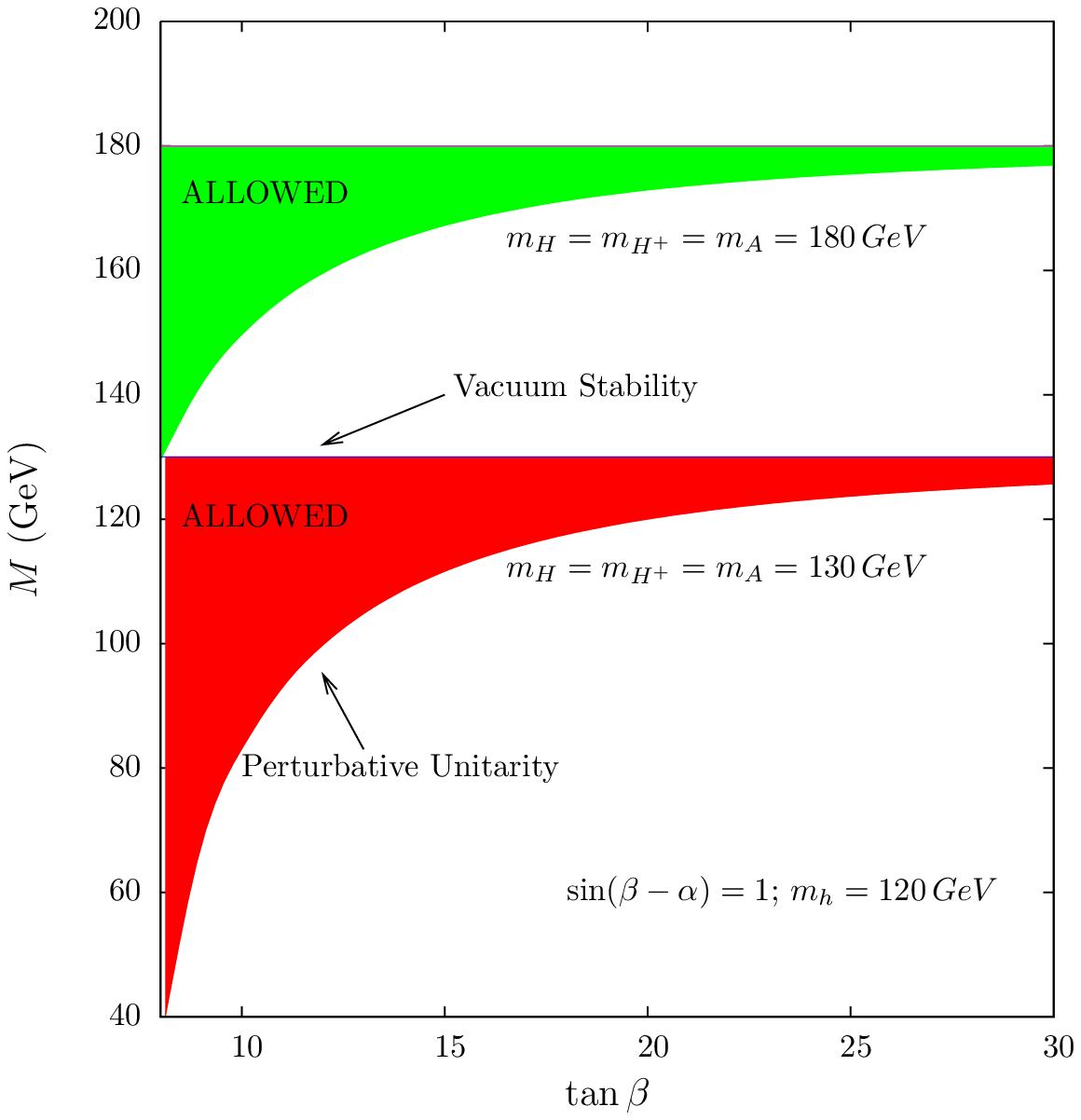}
\includegraphics[width=7.5cm]{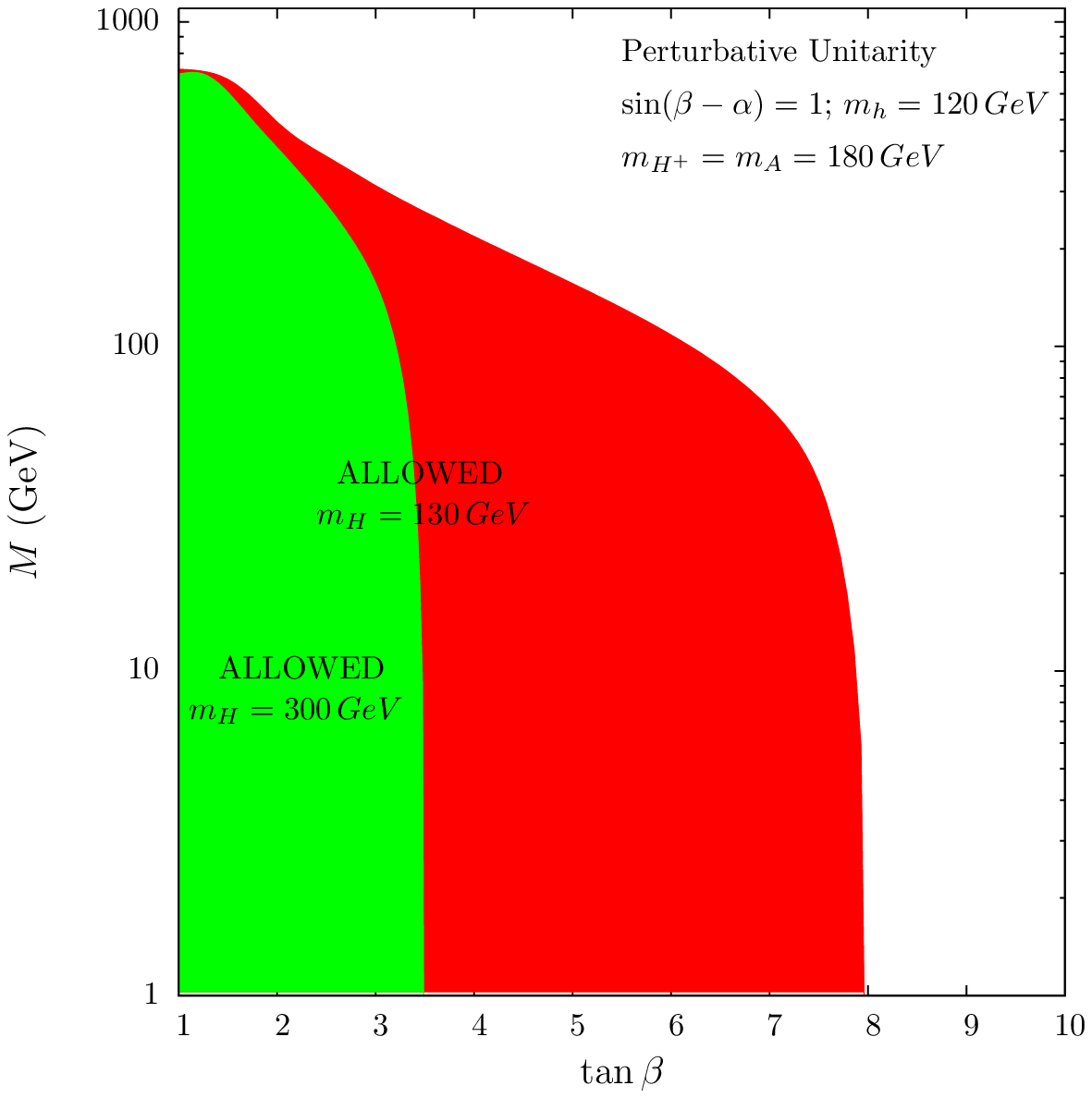}
\end{center}
%\vspace{-2cm}
\caption{Perturbative unitarity and vacuum stability limits for $M^2$
as a function of $\tan \beta$ with $\sin (\beta - \alpha)=1$. On the left panel
$M^2 > 0$ and on the right panel $M^2 < 0$.}
\label{fig:thlim1}
\end{figure}

\begin{figure}
\begin{center}
\includegraphics[width=7.4cm]{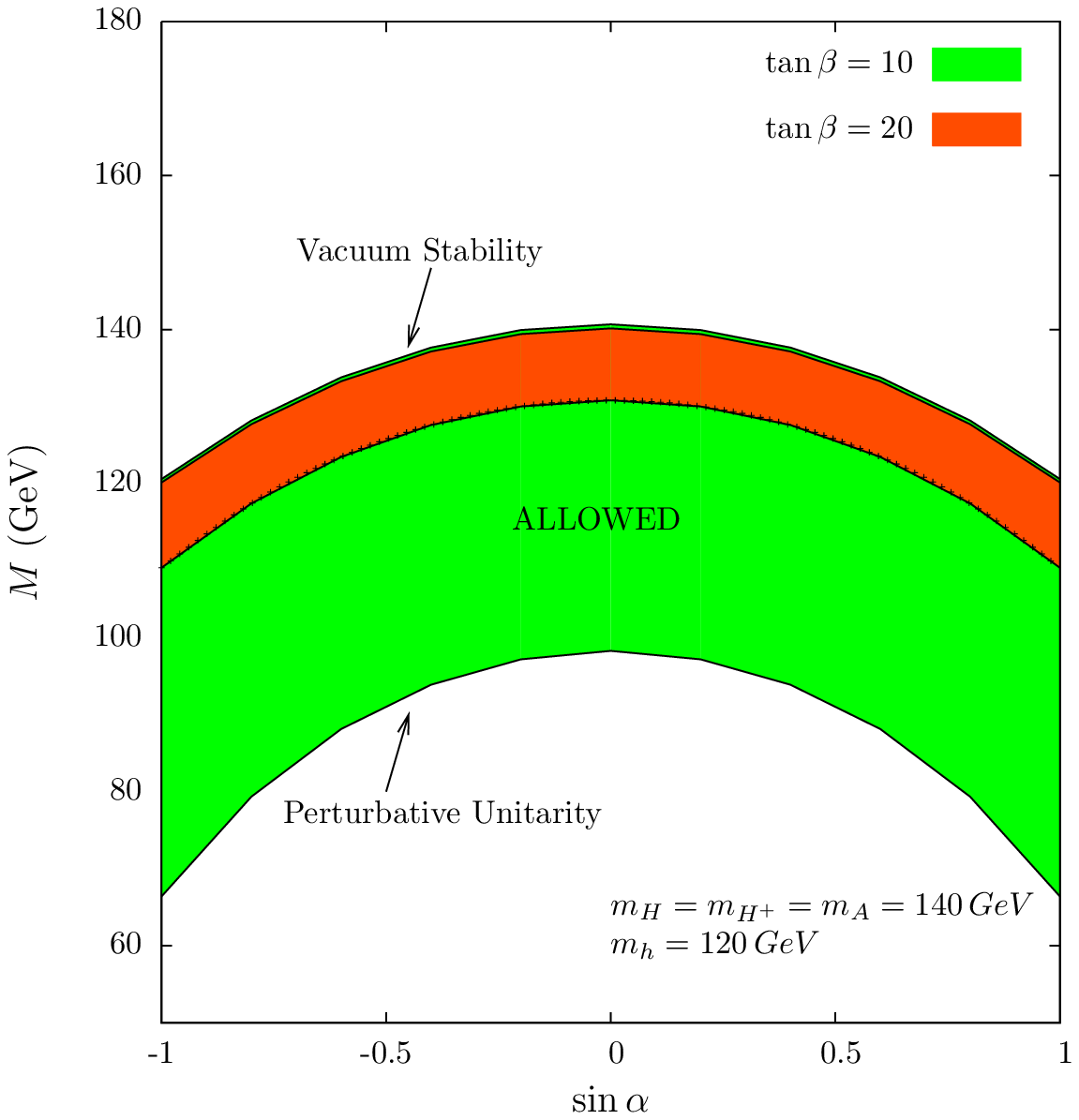}
\includegraphics[width=7.4cm]{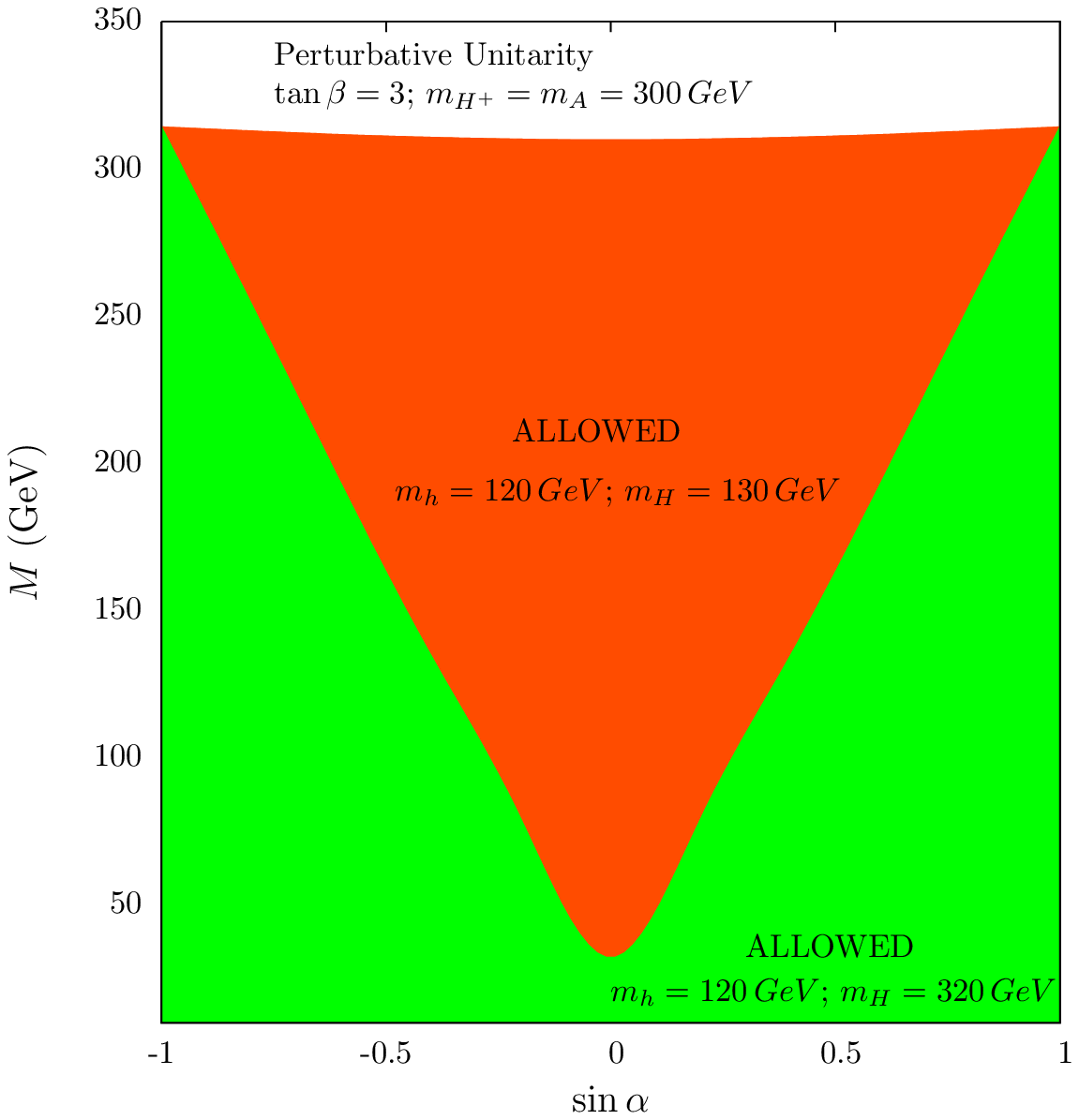}
\end{center}
%\vspace{-2cm}
\caption{Perturbative unitarity and vacuum stability limits for $M^2$
as a function of $\sin \alpha$ for several values of $\tan \beta$ and Higgs masses.
On the left panel
$M^2 > 0$ and on the right panel $M^2 < 0$.}
\label{fig:thlim2}
\end{figure}

In Fig.~\ref{fig:thlim1} we show how vacuum stability and perturbative unitarity constrain the parameter space of the model in the limit $\sin (\beta - \alpha) =1$. In the left panel, where $M^2 > 0$, it is clear that large values of $\tan \beta$ are allowed. However, if $\tan \beta$ is very large $M$ has to be very close and below the mass of the heaviest CP-even Higgs boson. Although very restrictive it should be noted that the MSSM lives in such a region (see for example the MSSM potential  in~\cite{GomezBock:2005if} where the comparison with the 2HDM potential is straightforward). In the right panel we consider the case $M^2 < 0$. Contrary to the previous scenario, now $M$ is much less constrained but $\tan \beta$ has to be rather small. We show the limits for $\tan \beta$ for two values of $m_H$ and the conclusion is that as the masses and/or $M$ grow, the allowed value of $\tan \beta$ becomes smaller. In Fig.~\ref{fig:thlim2} we plot the allowed region by the same theoretical constraints but now as a function of $\sin \alpha$. When $M^2 > 0$ (left panel) the allowed region, shown for $\tan \beta =10$ and $20$, has a mild dependence on $\sin \alpha$. When $M^2 < 0$ (right panel) there is a strong combined dependence on $m_H$, $m_h$ and $\sin \alpha$ for a fixed value of $\tan \beta$ and $M$ is forced to be very small for large values of the Higgs masses. Finally note that for $M^2 < 0$ the allowed region is hardly affected by the vacuum stability conditions.
%
%
%\begin{widetext}
%\end{widetext}
%
%\begin{widetext}
%\end{widetext}
%
%

There are several experimental constraints to consider. New contributions to the $\rho$ parameter stemming from Higgs states \cite{Rhoparam} have to comply with the current limits from precision measurements \cite{pdg}: $ |\delta\rho| \la 10^{-3}$. There are limiting cases though, related to an underlying custodial symmetry,  where the extra contributions to $\delta\rho$ vanish: $m_{H^\pm} \approx m_A$ or $m_{H^\pm} \approx m_h$ and $\sin (\beta - \alpha) = 1$ or $m_{H^\pm} \approx m_H$ and $\sin (\beta - \alpha) = 0$. Since we are only interested in the CP-even Higgs, only the constraints on $\tan \beta$ are of relevance to our study. These usually come associated with constraints on the charged Higgs mass. Values of $\tan \beta$ smaller than $\approx 1$ together with a charged Higgs with a mass below 100 $GeV$ are disallowed both by the constraints coming from $R_b$ (the $b$-jet fraction in $e^+e^-\to Z\to$ jets) \cite{LEPEWWG,SLD} and from $B_q \bar{B_q}$ mixing~\cite{Oslandk} for all Yukawa versions of the model. It has been shown in~\cite{bsgamma} that data from $B\to X_s \gamma$ impose a lower limit of $m_{H^\pm} \ga 290$\,GeV in models type II and III. This constraint no longer applies in the case of 2HDM type I and IV as discussed in~\cite{Aoki:2009ha}. From processes involving the tau lepton, only $\tau \to \mu \bar{\nu} \nu$~\cite{taumununu} constrains the 2HDM type IV. The analysis in~\cite{Aoki:2009ha} and~\cite{Logan:2009uf} show that these are very weak constraints on the leptonic version of the model especially because the LEP bound already excluded a charged Higgs below $\sim 100$ $GeV$~\cite{pdg}. There are no other constraints relevant to our analysis.

\section{Conclusions}
\label{sec:conclusions}

We have performed a detailed parton level analysis on the feasibility of the detection of a Higgs plus a high $p_T$ jet in the process $pp (gg+gq) \to h + jet \to \tau^+ \tau^- + jet$ at the LHC. We have considered both the leptonic and the semi-leptonic final states. Although a complete experimental analysis is needed, we have taken into account the effect of detector energy resolution which also defines non-instrumental effects of missing energy resolution. By taking into account these (main for our process) detector effects, which determines the resolution of the reconstructed Higgs boson peak, and which are necessary for a correct background estimation, we demonstrate that a 120--140 $GeV$ SM Higgs boson could be probed with less than 5 $fb^{-1}$ exploiting the suggested signature. For the same mass range a 5$\sigma$ discovery could be claimed with $\sim$ 30 $fb^{-1}$. Therefore, it is quite an appealing process for SM Higgs boson search along with others previously proposed in the literature. Moreover, the $pp \to h + jet \to \tau^+ \tau^- + jet$ process could become much more important in BSM models such as Supersymmetry or Technicolour or some specific 2HDMs. In all such scenarios the signal rate can be significantly higher than in the SM case due to an enhancement of the Higgs production or decay rate or both. Finally, we note that this process is the only one proposed so far for Higgs searches where only Yukawa couplings are directly involved.

We have then applied the results to models where the Higgs decay to leptons is enhanced relative to the SM. All models have in common the fact they have a specific type of 2HDM, the one where one doublet couples only to leptons while the other couples to all quarks, as a submodel. We investigated the different models that can be probed at the LHC and show that some regions of their parameter space are accessible with just a few $fb^{-1}$ of integrated luminosity.

\acknowledgments We thank Abdesslam Arhrib, Nuno Castro, Glen Cowan, Bruce Mellado, Alexander Pukhov, Junichi Tanaka,
Pedro Teixeira-Dias, Robert Thorne and Filipe Veloso for discussions. We thank Ricardo Gon\c{c}alo for his help with the trigger efficiencies.
We thank Johan Alwall, Michel Herquet and Fabio Maltoni for their help with MadGraph/MadEvent and Christian Schappacher
for providing us the vetocut.f routine for FormCalc.
RBG is supported by a Funda\c{c}\~ao para a Ci\^encia e Tecnologia Grant SFRH/BPD/47348/2008.
SM is financially supported in part by the scheme `Visiting Professor - Azione D - Atto Integrativo tra la
Regione Piemonte e gli Atenei Piemontesi. RS is supported by the FP7 via a Marie Curie Intra European Fellowship,
contract number PIEF-GA-2008-221707.

%%...............................................
\bibliographystyle{JHEP}

\bibliography{references}

\providecommand{\href}[2]{#2}\begingroup\raggedright\endgroup

\end{document}